\documentclass[pdflatex, sn-mathphys-ay]{sn-jnl}

\usepackage{graphicx}%
\usepackage{multirow}%
\usepackage{tikz-dependency}
\usepackage{tikz}
\usetikzlibrary{arrows.meta, positioning, automata}
\usepackage{amsmath,amssymb,amsfonts}%
\usepackage{amsthm}%
\usepackage{mathrsfs}%
\usepackage[title]{appendix}%
\usepackage{xcolor}%
\usepackage{textcomp}%
\usepackage{manyfoot}%
\usepackage{tcolorbox}
\usepackage{booktabs}%
\usepackage{algorithm}%
\usepackage{algorithmicx}%
\usepackage{subcaption}
\usepackage{algpseudocode}%
\usepackage{listings}%
\usepackage{pifont}
\usepackage{orcidlink}

\theoremstyle{thmstyleone}%
\theoremstyle{thmstyletwo}%

\theoremstyle{thmstylethree}%

\begin{document}

\title[Article Title]{Speaker-Aware Simulation Improves Conversational Speech Recognition}


\author*[1,2]{\fnm{Máté} \sur{Gedeon} \orcidlink{0009-0005-1429-8279}}\email{gedeonm@edu.bme.hu}

\author[2,3]{\fnm{Péter} \sur{Mihajlik} \orcidlink{0000-0001-7532-9773}}\email{mihajlik@tmit.bme.hu}

\affil[1]{\orgdiv{Dept. of Telecommunications and Artificial Intelligence}, \orgname{Budapest University of Technology and Economics}, \orgaddress{\city{Budapest}, \country{Hungary}}}

\affil[2]{\orgname{SpeechTex Ltd.}, \orgaddress{ \city{Budapest}, \country{Hungary}}}

\affil[3]{\orgname{ELTE Research Centre for Linguistics}, \orgaddress{ \city{Budapest}, \country{Hungary}}}


\abstract{Automatic speech recognition (ASR) for conversational speech remains challenging due to the limited availability of large-scale, well-annotated multi-speaker dialogue data and the complex temporal dynamics of natural interactions. Speaker-aware simulated conversations (SASC) offer an effective data augmentation strategy by transforming single-speaker recordings into realistic multi-speaker dialogues. However, prior work has primarily focused on English data, leaving questions about the applicability to lower-resource languages.
In this paper, we adapt and implement the SASC framework for Hungarian conversational ASR. We further propose C-SASC, an extended variant that incorporates pause modeling conditioned on utterance duration, enabling a more faithful representation of local temporal dependencies observed in human conversation while retaining the simplicity and efficiency of the original approach.
We generate synthetic Hungarian dialogues from the BEA-Large corpus and combine them with real conversational data for ASR training. Both SASC and C-SASC are evaluated extensively under a wide range of simulation configurations, using conversational statistics derived from CallHome, BEA-Dialogue, and GRASS corpora. Experimental results show that speaker-aware conversational simulation consistently improves recognition performance over naive concatenation-based augmentation. While the additional duration conditioning in C-SASC yields modest but systematic gains--most notably in character-level error rates--its effectiveness depends on the match between source conversational statistics and the target domain.
Overall, our findings confirm the robustness of speaker-aware conversational simulation for Hungarian ASR and highlight the benefits and limitations of increasingly detailed temporal modeling in synthetic dialogue generation.}
\keywords{Conversational speech, automatic speech recognition, simulated conversations, data augmentation}



\maketitle

\section{Introduction}\label{sec1}
Transcribing conversations in any language presents a challenging task for automatic speech recognition (ASR) systems. In everyday speech, people interrupt each other, speak at the same time, and hesitate \citep{Schegloff2000, Heldner2010}. Consequently, ASR models trained on clean, single-speaker recordings often fail to generalize to real multi-speaker dialogues with rapid turn-taking and overlapping speech \citep{Yu2016PIT}. Training robust conversational speech models requires large amounts of multi-speaker, dialogue-style audio \citep{CHiME-6}. Unfortunately, such richly annotated conversational data is scarce, especially for languages like Hungarian with fewer available resources. 

To address the scarcity of conversational speech data, researchers increasingly rely on synthetic conversation simulation, i.e. the generation of multi-speaker dialogues from isolated single-speaker utterances \citep{Yu2016PIT, SOT}. Over time, conversation simulation has evolved substantially: early approaches relied on simple utterance concatenation \citep{Fujita2019}, later methods incorporated timing statistics derived from real conversations \citep{Landini2022}, and more recent work has introduced speaker-aware simulated conversations (SASC), which explicitly model speaker-specific conversational behavior \citep{SASC}. While SASC has been shown to improve realism according to several intrinsic metrics, and a English simulated dataset has been released \citep{Libriconvo}, prior work has not demonstrated whether augmenting training data with SASC yields measurable improvements on real conversational ASR benchmarks. In this work, we close this gap through comprehensive experiments, providing the first evidence of the downstream utility of SASC on real conversational speech and its advantages compared to previous methods, and offer broader insights into the role of simulated conversations in ASR research.

Despite these advances, independence assumptions remain even in the SASC framework. In particular, the length of a speaker's next utterance does not influence the pause before or after it. In natural human interaction, however, such dependencies are well documented \citep{speech_planning}: longer planned turns are often preceded by longer gaps, whereas short reactive responses tend to follow with minimal delay. To address this limitation, we extend the SASC paradigm with a speaker- and duration-aware conversation simulation framework, termed C-SASC. The proposed approach jointly models (i) each speaker's characteristic timing behavior and (ii) the dependency between an utterance’s duration and the preceding pause. By dynamically sampling gap lengths conditioned on the upcoming utterance, C-SASC more closely reflects temporal patterns observed in natural dialogue.

We apply this framework to the Hungarian BEA-Large corpus \citep{bea_large}, which provides extensive single-speaker recordings across diverse speakers, making it well suited for conversation simulation. The generated dialogues are then combined with the BEA-Dialogue corpus \citep{bea_large} for ASR training and evaluation. These corpora are derived from the Hungarian BEA database \citep{bea2014, Mady2024RevisedAnnotation}, supplying benchmark for Hungarian ASR. Our experiments assess how these simulated conversations contribute to transcribing real Hungarian dialogues, highlighting the practical benefit of incorporating both speaker individuality and temporal dependency into simulation. Notably, this work represents the first application of SASC beyond the English language.

The main contributions of this work are as follows:
\begin{itemize}
    \item We adapt and apply the Speaker-Aware Simulated Conversations (SASC) framework to Hungarian conversational speech, demonstrating its applicability by performance improvement on a real speech corpus.
    \item We propose an extended variant of the framework, termed C-SASC, which incorporates utterance-duration-conditioned pause modeling while preserving the original simulation structure.
    \item We conduct a systematic investigation of statistics derived from three different conversational speech corpora (CallHome \citep{CallHome}, BEA-Dialogue \citep{bea_large}, and GRASS \citep{grass}), enabling cross-corpus comparisons.
    \item We provide a comprehensive evaluation of both SASC and C-SASC under multiple simulation configurations, including different speaker-pairing strategies and corpus-dependent statistics.
    \item We analyze the effect of simulated dataset size on downstream automatic speech recognition performance, highlighting scaling behavior and diminishing returns.
    \item We examine the impact of room impulse response (RIR) augmentation within speaker-aware simulated dialogues and identify conditions under which it may be beneficial or detrimental.
\end{itemize}

The remainder of this paper is organized as follows. Section 2 reviews related work. Section 3 presents the theoretical framework of the proposed methodology and describes its application to dataset generation. Section 4 outlines the experimental setup and reports the results. Section 5 discusses the findings, and Section 6 concludes the paper with a summary of the main contributions and directions for future research.

\section{Related Work}
The generation of synthetic or simulated conversational data has become an increasingly vital strategy in both automatic speech recognition (ASR) and speaker diarization, particularly in scenarios or languages where real conversational recordings are scarce or prohibitively expensive to obtain. The main objective of these approaches is to solve the limited availability of annotated multi-speaker interactions using prevalent single-speaker speech data, by simulating dialogues from them that approximate real conversational dynamics.

Some efforts in data augmentation primarily target single-speaker scenarios in low-resource languages, aiming to expand datasets using available resources. For instance, \citet{bartelds2023} demonstrated that self-training and text-to-speech (TTS) augmentation can substantially improve ASR performance in minority and regional languages. Their results showed that generating synthetic speech from text or pseudo-labels can yield relative word error rate (WER) reductions of up to 25\%. While such techniques effectively increase data volume, they remain confined to single-speaker conditions and thus cannot capture the overlapping speech and turn-taking phenomena that characterize natural dialogues.

In contrast, a parallel line of research has focused on simulation techniques that explicitly model the temporal, overlapping, and alternation patterns observed in real conversations. Early approaches to synthetic conversation generation were quite simplistic: for example, one could randomly concatenate isolated utterances from different speakers and insert silences between them \citep{Fujita2019, Yamashita2022Naturalness}. While easy to implement, this yielded unnatural dialogue -- the timing of responses and overlaps did not resemble real human interactions. Subsequent methods became more sophisticated by statistically modeling the gaps and overlaps between turns by sampling pause durations and overlap lengths based on distributions observed in real dialogue corpora \citep{Landini2022}, rather than using arbitrary or fixed gaps. This statistical turn-taking simulation was a significant improvement: it produced more lifelike timing and even led to better speaker diarization accuracy in benchmarks \citep{Landini2022MultiSpeakerEEND}, compared to the naive random concatenation approach. However, these improved simulation techniques still had a notable limitation as they treated all speakers as if they were the same. This came from the fact that the timing patterns were modeled by an overall distribution, ignoring each individual's speaking style. Real conversations are not one-size-fits-all; each speaker has personal habits (some consistently pause longer than others, for instance). By relying on general timing distributions, these simulations missed these nuances and thus failed to capture some of the more complex, natural dynamics of multi-speaker exchange.

To address this issue, the concept of speaker-aware simulated conversations (SASC) has emerged \citep{SASC}, which explicitly incorporates speaker-specific behavior into the generation of synthetic dialogues. The key idea in SASC is that every speaker in a simulated conversation should retain their own characteristic timing patterns -- much like they would in a real interaction. Rather than using a single general distribution for all pause lengths or overlap probabilities, SASC learns per-speaker timing distributions from data. For example, if one person tends to have shorter gaps on average before they respond, the simulation will reflect that consistently whenever that same virtual speaker takes a turn. In practice, the SASC framework models turn-taking via a Markov chain model, determining the sequence of speaker turns in a probabilistic but realistic way. 

Evaluations on English corpora showed that the speaker-aware approach aligned more closely with real conversational patterns in several intrinsic metrics \citep{SASC}. These results demonstrated that modeling each speaker's timing consistency can substantially improve the realism of synthetic multi-speaker data.

The applicability of this approach was demonstrated by LibriConvo \citep{Libriconvo}, created by taking the LibriTTS corpus \citep{LibriTTS} (comprised of many individual readers) and transforming it into multi-speaker dialogues. To make the conversations sound more plausible, the LibriConvo pipeline ensured topical consistency in each dialogue, with grouping utterances by book, so that when they were interleaved as a conversation, they appeared to be discussing the same topic or narrative. Additionally, LibriConvo applied heuristic acoustic simulations -- adding reverberation with a room impulse response model -- so that all speakers in a simulated conversation sounded like they were in the same physical room. The result of this construction was a synthetic dataset of over 240 hours of speech, spanning 1,496 dialogues and 830 unique speakers.

A complementary advance in multi-talker ASR simulation was introduced by \citet{yang2023} with a data-driven method for generating realistic overlapping speech. Rather than inserting overlaps randomly, they represented overlap structures as discrete tokens learned from real conversational data using a statistical language model. This enabled synthetic mixtures to exhibit more natural overlap sequences, resulting in measurable WER improvements across multiple datasets. The idea of learning overlap dynamics from authentic conversational data closely parallels the motivation behind the SASC framework, although their approach focused specifically on overlap patterns rather than complete conversational flow or speaker individuality.

A notable predecessor to LibriConvo is LibriHeavyMix \citep{libriheavymix}, a 20,000-hour dataset designed to advance research in single-channel speech separation, ASR, and speaker diarization under far-field, overlapping conditions. Built from the LibriSpeech corpus, LibriHeavyMix employed an earlier generation method of conversational simulation without explicit modeling of speaker individuality or fine-grained temporal dependencies. The resulting corpus provides a reproducible pipeline for joint evaluation of speech separation, recognition, and diarization models. Although LibriHeavyMix achieved impressive scale, its conversational realism was limited by the absence of speaker-aware or turn-taking models. This limitation motivated subsequent work such as SASC and LibriConvo, which shifted focus toward statistically grounded and speaker-conditioned conversation simulation to capture more human-like dialogue behavior.

Further progress in end-to-end neural diarization (EEND) has underscored the importance of natural conversational simulation. \citet{Yamashita2022Naturalness} proposed a structured approach to generating synthetic conversations by defining four distinct types of speaker transitions and arranging them sequentially to reflect real dialogue dynamics. The resulting datasets exhibited silence and overlap ratios statistically similar to real corpora and led to improved diarization accuracy on benchmark datasets such as CallHome and CSJ (Corpus of Spontaneous Japanese). This work reinforced the view that the realism of simulated data--particularly in turn-taking and overlap structure--directly influences model performance.

Overall, these studies illustrate a clear evolution from basic data augmentation toward behaviorally grounded, probabilistic, and structure-aware conversation simulation. Despite this progress, significant gaps remain, particularly in extending these simulation techniques to underrepresented languages such as Hungarian, offering promising directions for ongoing research.

\section{Methodology}
\subsection{SASC}
To provide an overall understanding of the underlying algorithm, we briefly summarize the speaker-aware simulated conversation (SASC) framework. For a complete and detailed description, we refer the reader to \citet{SASC}.

Pauses and overlaps at speaker transitions are characterized by a variable $\delta$, where $\delta < 0$ denotes overlap and $\delta \geq 0$ denotes a pause. The integral over the negative domain of the distribution corresponds to the probability of overlap, $p_{\text{overlap}}$ (see Fig.~\ref{fig:pause_distributions} for an example).

Kernel density estimation (KDE) is applied to obtain continuous estimates of pause-length distributions. To ensure temporal consistency, two distributions are defined: $\hat{D}_{=}$ models the average pause lengths (per speaker) for same-speaker transitions (i.e., no speaker change), while $\hat{D}_{\neq}$ models the average pause lengths (per speaker) for different-speaker transitions. For each speaker $s$, an initial base value $\mu$ is sampled from the corresponding distribution, and subsequent pauses are generated by adding a deviation term $v$:
\[
\delta_n =
\begin{cases}
\mu^{\text{same}}_s + v, & \text{if } X_n = X_{n-1},\; v \sim V_{=},\\[4pt]
\mu^{\text{diff}}_s + v, & \text{if } X_n \neq X_{n-1},\; v \sim V_{\neq}.
\end{cases}
\]
Here, $X_n$ denotes the speaker of the $n$-th utterance. The distributions $V_{=}$ and $V_{\neq}$, referred to as \emph{deviation distributions}, capture the variability of pauses around a speaker-specific mean and are also estimated from data.

Speaker turn-taking is modeled using a first-order Markov chain (extendable to higher-order), where the transition matrix $P_{\mathrm{turn}}$ defines the probability of the next speaker conditioned on the previous one.

All speakers are placed within a shared acoustic environment: a single room is selected from the available room impulse responses (RIRs), and each speaker is assigned a unique spatial position within that room.

A compact description of the overall process is given in the pseudocode shown in Algorithm~\ref{sasc_algo}.

\begin{algorithm}[h]
\caption{\label{sasc_algo} Simplified speaker-aware simulation conversation}
\begin{algorithmic}[1]
\State Select $N_{\mathrm{spk}}$ speakers ($\mathcal{S}'$); assign RIRs (same room, different positions)
\State Select initial speaker $X_1$
\For{$n = 1 \dots N_{\mathrm{u}}$}
  \If{$n > 1$}
    \State $X_n \sim P_{\mathrm{turn}}(X_{n-1})$
  \EndIf
  \State Sample utterance $u_n \in U_{X_n}$ and apply RIR to obtain $y_n$
  \If{$n = 1$}
    \State $\delta = 0$
  \ElsIf{$X_n = X_{n-1}$}
    \If{$X_n$ first pause}
      \State $\mu^{\text{same}}_s \sim \hat{D}_{=}$
    \EndIf
    \State $\delta = \mu^{\text{same}}_s + v,\; v \sim V_{=}$
  \Else
    \If{$X_n$ first pause}
      \State $\mu^{\text{diff}}_s \sim \hat{D}_{\neq}$
    \EndIf
    \State $\delta = \mu^{\text{diff}}_s + v,\; v \sim V_{\neq}$
  \EndIf
  \State Append $y_n$ to the dialogue with pause $\delta$
\EndFor
\end{algorithmic}
\end{algorithm}

\subsection{C-SASC}
\label{sec:sasc_plus}
The original SASC framework generates pauses using speaker-dependent distributions that remain fixed throughout a dialogue. While this formulation captures differences between speakers and transition types, it implicitly assumes that pause behavior is independent of utterance length.

C-SASC relaxes this assumption by allowing pause behavior to vary as a function of utterance duration, thereby providing a lightweight mechanism to account for dependencies not modeled in SASC. The motivation for this comes from previous research suggesting that the pause before an utterance is influenced by the length of that utterance \citep{speech_planning}. By conditioning pause deviations on duration, C-SASC can generate inter-utterance gaps that vary systematically with the length of surrounding utterances, while requiring only minimal modifications to the inference procedure.

As in the original SASC formulation, pauses in C-SASC are generated as the sum of a speaker-dependent base value and a residual term. The key extension is that the residual term is explicitly conditioned on utterance duration. For each speaker \( s \) and transition type (\(=\) for same-speaker continuation and \( \neq \) for speaker change), pauses are generated according to
\[
\delta_n =
\begin{cases}
\mu^{\text{same}}_s + r_n, & \text{if } X_n = X_{n-1}, r_n \sim V_{=}(r \mid d_n) \\[4pt]
\mu^{\text{diff}}_s + r_n, & \text{if } X_n \neq X_{n-1}, r_n \sim V_{\neq}(r \mid d_n)
\end{cases}
\]
where \( X_n \) denotes the speaker of the \(n\)-th utterance and \( \delta_n \) is the inter-utterance gap preceding it. The terms \( \mu^{\text{same}}_s \) and \( \mu^{\text{diff}}_s \) represent speaker-specific base pause values for same-speaker continuation and speaker change, respectively, and are sampled once per dialogue in the same way as in SASC.

The residual term \( r_n \) captures local temporal variability around the speaker-specific baseline and is drawn from a duration-conditioned distribution. Let \( d_n \) denote the duration of the utterance following the gap \( \delta_n \). For each transition type \( \tau \in \{=,\neq\} \), we model the conditional residual density \( p_\tau(r \mid d) \) using a kernel conditional density estimator with separate kernels over the residual and duration dimensions.

Given a set of training samples \( \{(r_i, d_i)\}_{i=1}^{N_\tau} \) extracted from the corresponding transition type \( \tau \), the conditional density is estimated as
\[
\hat{p}_\tau(r \mid d^\ast)
=
\frac{
\sum_{i=1}^{N_\tau}
K_r\!\left(\frac{r - r_i}{h_r}\right)
K_d\!\left(\frac{d^\ast - d_i}{h_d}\right)
}{
\sum_{i=1}^{N_\tau}
K_d\!\left(\frac{d^\ast - d_i}{h_d}\right)
},
\]
where \( K_r(\cdot) \) and \( K_d(\cdot) \) are kernel functions (Gaussian in our experiments), \( h_r \) and \( h_d \) are the corresponding bandwidth parameters. This formulation is based on the one reported in \citet{holmes2012}, and is named Nadaraya-Watson
conditional density estimator \citep{nadaraya, watson}, in which the conditional density over residuals is obtained by locally weighting samples according to the proximity of their associated durations.

The mean pause distributions are modeled in the same way as in SASC. For each transition type \( \tau \in \{=,\neq\} \), we collect the per-speaker mean pauses \( \{\mu^{\tau}_s\}_{s=1}^{S} \) and estimate their distribution using one-dimensional kernel density estimation:
\[
\hat{p}_{\tau}(\mu)
=
\frac{1}{S h_{\mu}}
\sum_{s=1}^{S}
K\!\left(
\frac{\mu - \mu^{\tau}_s}{h_{\mu}}
\right),
\]
where \( K(\cdot) \) is a Gaussian kernel and \( h_{\mu} \) is the bandwidth. The resulting distributions, denoted by \( \hat{D}_{=} \) and \( \hat{D}_{\neq} \), are used to sample speaker-specific base pause values at dialogue initialization.

At inference time, a speaker-specific baseline pause \( \mu^{\tau}_s \) is sampled once per dialogue from the corresponding distribution \( \hat{D}_{\tau} \) and remains fixed throughout the dialogue. This value represents the characteristic timing tendency of speaker \( s \) under the given transition type. In contrast, local variability is introduced via the duration-conditioned residual term, allowing pause behavior to adapt dynamically to utterance duration.

To mitigate distribution skewness and improve Gaussianity while preserving support over both negative (overlap) and positive (pause) values, a Yeo--Johnson power transformation \citep{Yeo-Johnson} is applied prior to KDE estimation.

In the original SASC formulation, no explicit bandwidth selection strategy was specified for kernel density estimation. In C-SASC, this design choice is made explicit to accommodate the increased model complexity introduced by the conditionality.
For the one-dimensional KDEs modeling speaker-specific baseline pauses, Silverman’s rule of thumb \cite{silverman1986} is used to initialize the bandwidth. Given per-speaker mean pauses \( \mu = \{\mu^{\tau}_s\}_{s=1}^{S} \) for transition type \( \tau \in \{=,\neq\} \), the bandwidth (h) is computed as
\[
h_{\mu} = 0.9 \, \min\!\left(\sigma_{\mu}, \frac{\mathrm{IQR}_{\mu}}{1.34}\right) S^{-1/5},
\]
where \( \sigma_{\mu} \) and \( \mathrm{IQR}_{\mu} \) denote the standard deviation and interquartile range of the sample, respectively.

For the conditional KDEs used to model duration-dependent residuals, separate bandwidths are estimated for the residual and duration dimensions using Scott’s rule \citep{scott1992}. Let \( \{(r_i, d_i)\}_{i=1}^N \) denote the training samples for a given transition type. The bandwidths are given by
\[
h_r = \sigma_r \, N^{-1/6}, \qquad
h_d = \sigma_d \, N^{-1/6},
\]
where \( \sigma_r \) and \( \sigma_d \) are the empirical standard deviations of the residual and duration samples, respectively. To ensure numerical stability and to prevent excessively small kernels in low-density regions of the duration space, both bandwidths can be lower-bounded as
\[
h_r \leftarrow \max(h_r, \varepsilon_r), \qquad
h_d \leftarrow \max(h_d, \varepsilon_d),
\]
with \( \varepsilon_r \) and \( \varepsilon_d \) denoting small positive constants.

With this combination of Silverman’s and Scott’s heuristics we aim to provide a framework, which is adaptive to the data at hand. Algorithm~\ref{sascplus_algo}. summarizes the updated pseudocode.

\begin{algorithm}[h]
\caption{\label{sascplus_algo} C-SASC pseudocode}
\begin{algorithmic}[1]
\State Select $N_{\mathrm{spk}}$ speakers ($\mathcal{S}'$); assign RIRs (same room, different positions)
\State Select initial speaker $X_1$
\For{$n = 1 \dots N_{\mathrm{u}}$}
  \If{$n > 1$}
    \State $X_n \sim P_{\mathrm{turn}}(X_{n-1})$
  \EndIf
  \State Sample utterance $u_n \in U_{X_n}$ with duration $d_n$ and apply RIR to obtain $y_n$
  \If{$n = 1$}
    \State $\delta_n \gets 0$
  \ElsIf{$X_n = X_{n-1}$}
    \If{first same-speaker pause for $X_n$}
      \State $\mu^{\text{same}}_s \sim \hat{D}_{=}$
    \EndIf
    \State $r_n \sim V_{=}(r \mid d_n)$
    \State $\delta_n \gets \mu^{\text{same}}_s + r_n$
  \Else
    \If{first different-speaker pause for $X_n$}
      \State $\mu^{\text{diff}}_s \sim \hat{D}_{\neq}$
    \EndIf
    \State $r_n \sim V_{\neq}(r \mid d_n)$
    \State $\delta_n \gets \mu^{\text{diff}}_s + r_n$
  \EndIf
  \State Append $y_n$ to the dialogue with inter-utterance gap $\delta_n$
\EndFor
\end{algorithmic}
\end{algorithm}

\subsection{Extracting statistics from real speech corpora}
Speaker-aware conversation simulation relies on three core statistical components: (i) pauses between consecutive utterances by the same speaker, (ii) pauses at speaker changes, and (iii) speaker transition probabilities.  

In the proposed C-SASC framework, (i) and (ii) are additionally conditioned on utterance duration. In practice, all statistics were extracted from three conversational speech corpora: the English CallHome corpus~\citep{CallHome}, the training set of the Hungarian BEA-Dialogue~\citep{bea_large}, and the Austrian German GRASS corpus~\citep{grass}. This selection enables comparisons of statistics not only across domains but also across languages.

The CallHome corpus consists of English telephone conversations recorded in stereo format, with each speaker captured on a separate channel. This setup allows for straightforward extraction of speaker turns, utterance durations, and inter-utterance gaps. During the construction of an English simulated conversational dataset \citep{Libriconvo} however, we observed that the publicly available annotations\footnote{\url{https://huggingface.co/datasets/talkbank/callhome}} lack sufficient temporal precision. Since both pause durations and utterance lengths are central variables in (C-)SASC, this limitation renders the original annotations unsuitable for our purposes.

To obtain reliable temporal information, the CallHome recordings were re-annotated using an automatic voice activity detection (VAD) system. We employed the Silero VAD model \citep{Silero} to derive accurate speech activity segments independently for each channel. From these segments, we extracted utterance durations as well as the pauses between consecutive utterances. These form the empirical basis for estimating both marginal pause distributions and duration-conditioned residual distributions used in C-SASC.

To complement the English data with statistics derived from Hungarian conversational speech, we also utilized the training portion of the BEA-Dialogue corpus. As with CallHome, we identified inaccuracies in the provided temporal annotations. In particular, word- or unit-level timestamps frequently touched, implying zero-length pauses between speakers, whereas auditory inspection revealed clearly audible gaps. Because such errors directly affect both pause statistics and utterance duration estimates, a reprocessing of the annotations was required.

Unlike CallHome, BEA-Dialogue recordings are stored as single-channel audio containing all speakers, which precludes direct speaker-wise VAD segmentation. Consequently, a multi-step procedure was adopted. First, word-level alignments were obtained using the Montreal Forced Aligner (MFA) \citep{mfa}, based on the available transcripts. These alignments provide temporal boundaries for individual words but do not directly yield utterance-level segments. Since C-SASC operates on utterance-level durations, we reconstructed sentence-like segments by applying an automatic punctuation restoration model\footnote{\url{https://huggingface.co/oliverguhr/fullstop-punctuation-multilang-large}} to the transcripts. The resulting sentence boundaries were then mapped back to the aligned audio, enabling estimation of utterance durations and the pauses between them.

As a third data source, we incorporated the GRASS corpus, which contains conversational speech with reliable speaker labeling and precise timing information. This makes it suitable for extracting both pause statistics and utterance duration distributions, without any processing. Including GRASS broadens the empirical basis of C-SASC by introducing additional conversational styles and recording conditions, thereby reducing corpus-specific bias in both marginal and duration-conditioned models.

For all three corpora, pause statistics were modeled using kernel density estimation (KDE) with a Gaussian kernel, yielding smooth, continuous probability density functions suitable for sampling during simulation. For SASC, we retained the default bandwidth parameter $\alpha = 0.1$ used in prior work. For C-SASC, we employed the bandwidth selection heuristics introduced in \textit{Section 3.2}.

\begin{figure}[h]
    \centering
    \begin{subfigure}{0.48\linewidth}
        \centering
        \includegraphics[width=\linewidth]{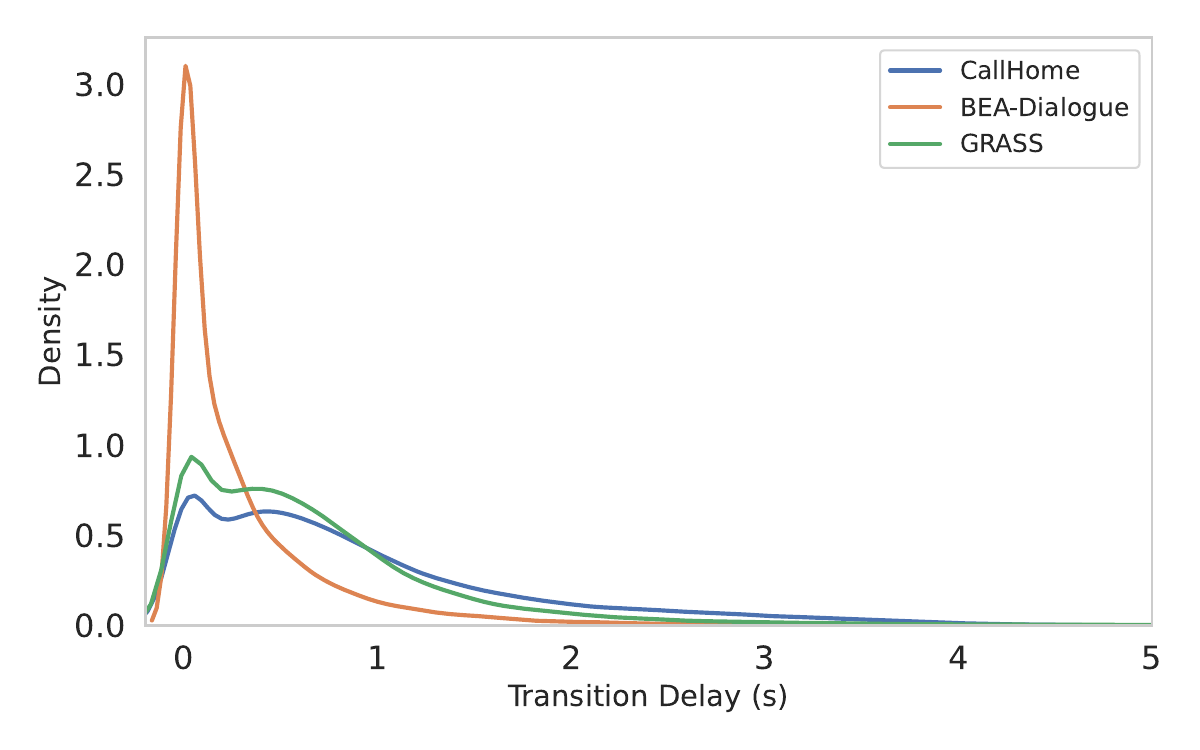}
        \caption{Same speaker (SASC)}
        \label{fig:same_spk_sasc}
    \end{subfigure}
    \hfill
    \begin{subfigure}{0.48\linewidth}
        \centering
        \includegraphics[width=\linewidth]{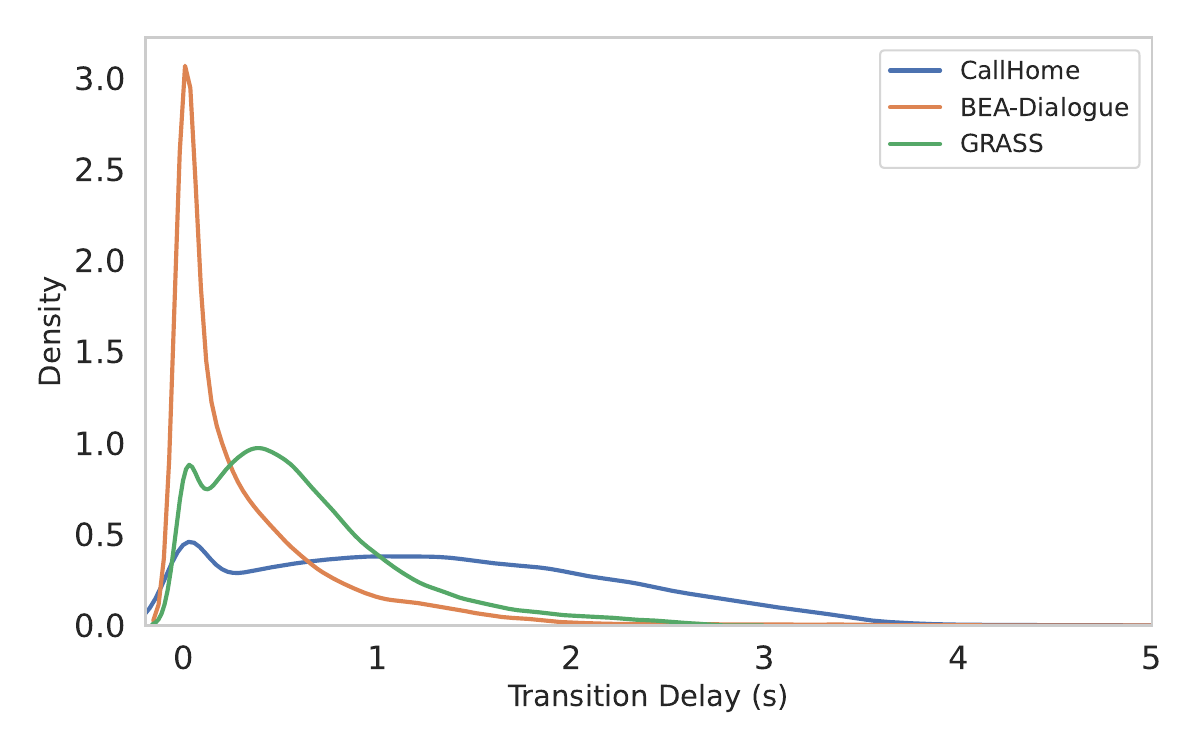}
        \caption{Same speaker (C-SASC)}
        \label{fig:same_spk_sasc_plus}
    \end{subfigure}
    \vspace{0.5em}
    \begin{subfigure}{0.48\linewidth}
        \centering
        \includegraphics[width=\linewidth]{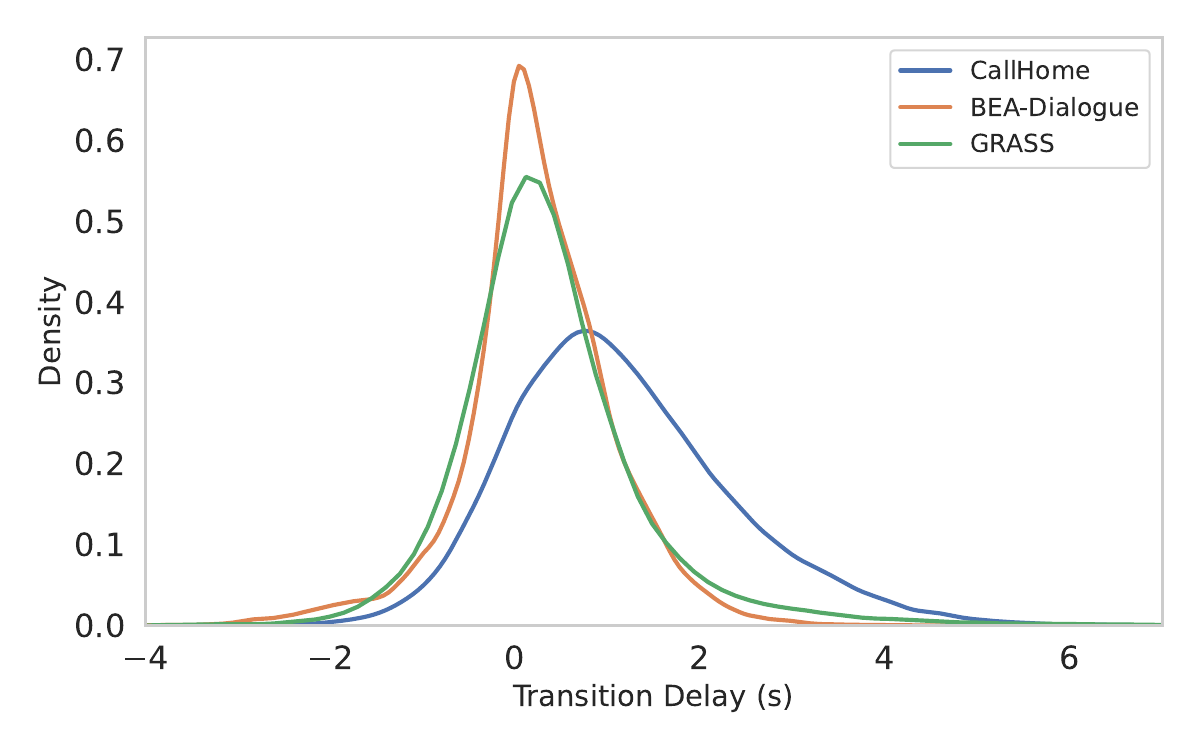}
        \caption{Different speakers (SASC)}
        \label{fig:diff_spk_sasc}
    \end{subfigure}
    \hfill
    \begin{subfigure}{0.48\linewidth}
        \centering
        \includegraphics[width=\linewidth]{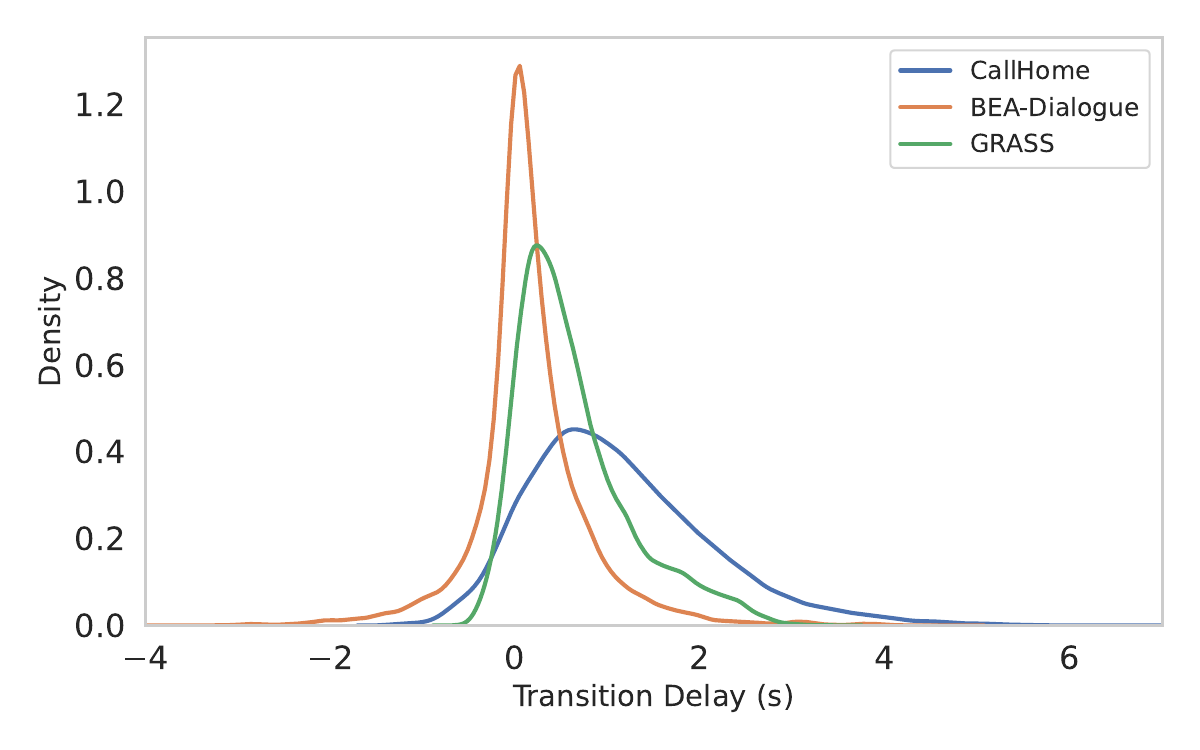}
        \caption{Different speakers (C-SASC)}
        \label{fig:diff_spk_sasc_plus}
    \end{subfigure}
    \caption{Modeled pause distributions for consecutive utterances under different speaker conditions.}
    \label{fig:pause_distributions}
\end{figure}

Since static visualization of the conditional distributions used in C-SASC is challenging, we instead analyze the resulting gap-length distributions after simulation. That is, we consider the \emph{posterior} distributions produced by the generative process, rather than the \emph{a priori} mean and deviation components. Marginal pause distributions for consecutive utterances by the same speaker are shown in Fig.~\ref{fig:same_spk_sasc} and Fig.~\ref{fig:same_spk_sasc_plus}, while Fig.~\ref{fig:diff_spk_sasc} and Fig.~\ref{fig:diff_spk_sasc_plus} illustrate the corresponding distributions for speaker changes. Negative values correspond to overlapping speech for both SASC and C-SASC.

\begin{figure}[ht] 
    \centering 
    \includegraphics[width=0.75\linewidth]{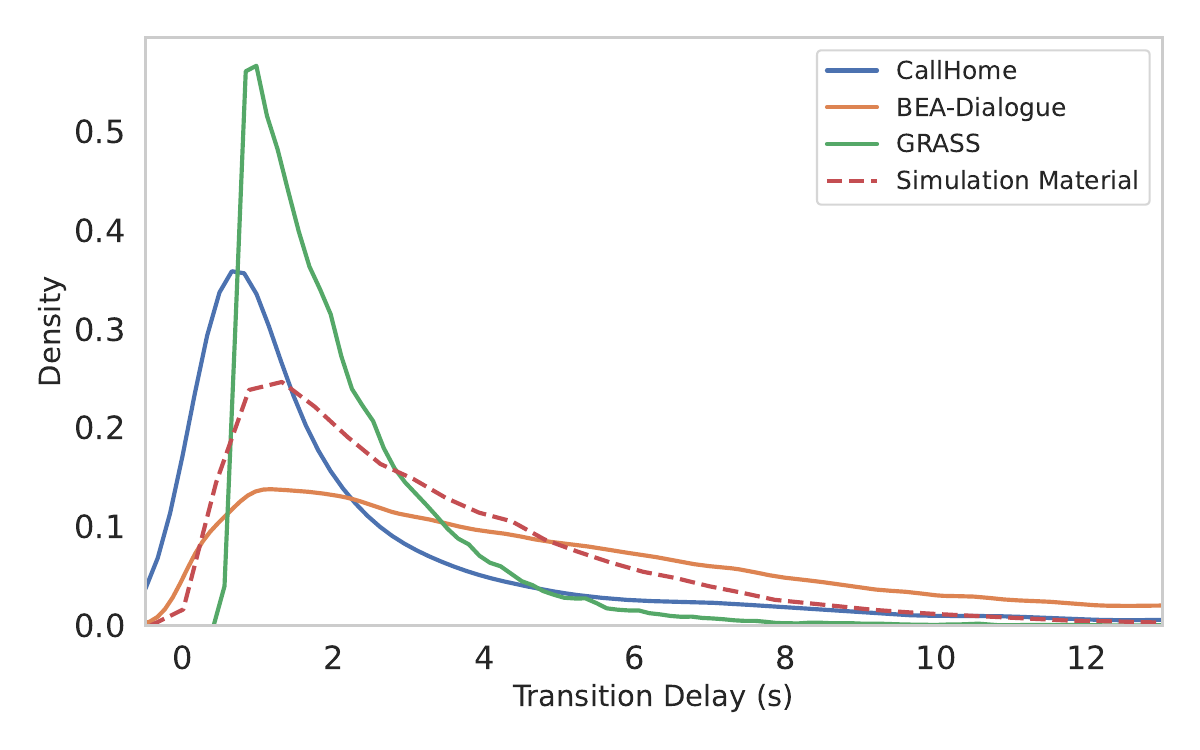} 
    \caption{Utterance duration distributions across corpora.} 
    \label{fig:utterance_length} 
\end{figure}

The differences between the curves produced by SASC and C-SASC for the same corpus arise primarily from differences in the underlying utterance duration distributions. Figure~\ref{fig:utterance_length} illustrates these duration distributions, together with the distribution of utterances used during simulation, on which the conditional modeling of C-SASC relies. Larger mismatches between training and simulation duration distributions lead to more pronounced differences between the two methods.

\begin{figure}[ht]
    \centering
    \begin{tikzpicture}[
        ->, >=Stealth,
        node distance=4cm,
        thick,
        state/.style={circle, draw, minimum size=1cm, font=\sffamily\bfseries, align=center},
        edge label/.style={font=\sffamily\small}
    ]

    \node[state] (A) {A};
    \node[state, right=of A] (B) {B};

    \draw[bend left] (A) to node[edge label, above]{0.633} (B);
    \draw[bend left] (B) to node[edge label, below]{0.631} (A);
    \draw (A) edge[loop above] node[edge label]{0.367} (A);
    \draw (B) edge[loop above] node[edge label]{0.369} (B);

    \end{tikzpicture}
    \caption{Speaker transition model estimated from the CallHome corpus.}
    \label{fig:fsa_ab}
\end{figure}
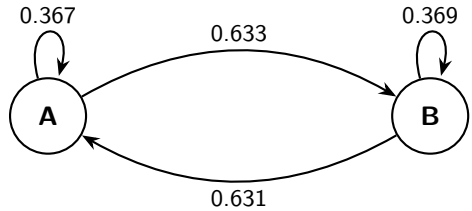

Speaker transition probabilities were estimated from the CallHome corpus using a first-order Markov chain. The resulting transition model, illustrated in Fig.~\ref{fig:fsa_ab}, governs the likelihood of speaker continuation versus speaker change independently of pause modeling. This transition model was used unchanged across all experiments, regardless of the corpus from which pause statistics were derived.

\subsection{Creating the simulated dataset(s)}
Simulated dialogues were generated using recordings from speakers in the BEA-Large corpus who do not appear in the BEA-Dialogue validation or evaluation sets, with each simulated conversation consisting of two speakers. To control speaker reuse across dialogues, we introduced a parameter limiting the number of distinct speaker pairs in which a given speaker may appear, ranging from one to five.

To obtain utterance units comparable to sentence-level segments and suitable for dialogue simulation, recordings were filtered by duration, retaining only segments between 2 and 10 seconds. The durations are explicitly used in C-SASC to condition pause generation. For each speaker pair, the longest possible simulated dialogue was constructed. This was achieved by using a Markov chain to iteratively determine the sequence of speaker turns and, consequently, the number of utterances required from each speaker for a dialogue consisting of \( n \) segments. The value of \( n \) was incremented until the maximum dialogue length was reached for which sufficient recordings were available from both speakers. The original chronological ordering of each speaker’s recordings was preserved to maintain speaker-wise coherence.

The following infobox summarizes basic statistics of the simulated dataset under the setting where each speaker participates in only a single pair. These statistics can be straightforwardly extrapolated to scenarios in which speakers appear in multiple pairs.

\begin{center}
\begin{tcolorbox}[
    colback=gray!5,
    colframe=gray!60,
    boxrule=0.6pt,
    arc=2mm,
    width=0.8\linewidth,
    title=\textbf{Statistics of the "unit" simulated dataset},
    fonttitle=\bfseries\sffamily,
    coltitle=black,
    center title
]
\begin{tabular}{l r}
Number of dialogues & 120 \\
Number of speakers & 240 \\
Average number of utterances (per dialogue) & 364.28 \\
Average utterance duration (per dialogue) & 4.38 s \\
Average dialogue length & 1595.53 s \\
\end{tabular}
\end{tcolorbox}
\end{center}

\begin{figure}[ht] 
    \centering 
    \includegraphics[width=0.75\linewidth]{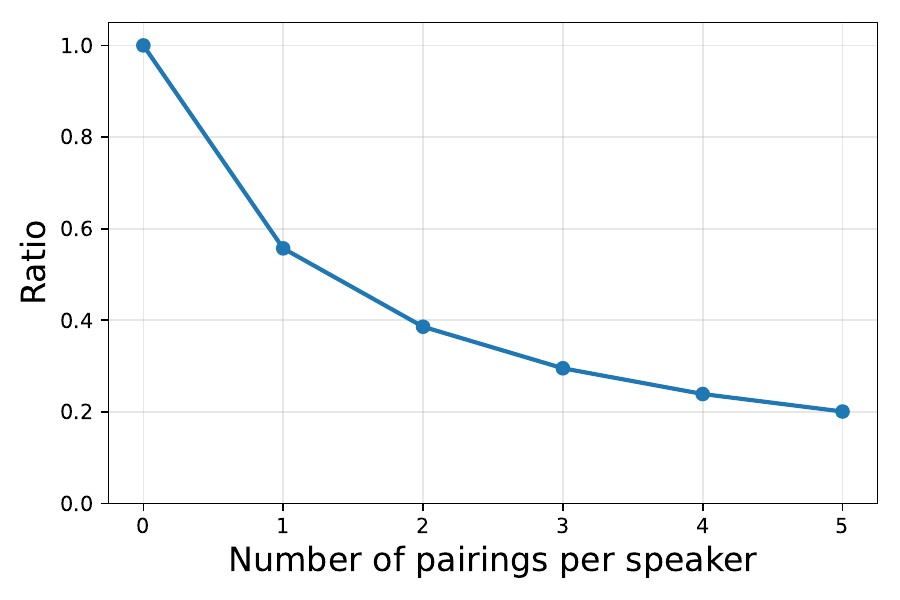} 
    \caption{Ratio of the BEA-Dialogue training set to the whole training set with increasing simulation size.} 
    \label{fig:ratio} 
\end{figure}

Figure~\ref{fig:ratio} illustrates the ratio between the BEA-Dialogue training set and the simulated dataset as a function of the number of speaker pairs in which each speaker appears. As this parameter increases, the size of the simulated dataset grows proportionally, enabling controlled scaling of the training data.

\section{Experiments}
\subsection{Experiment setup}
During training and evaluation, we followed the methodology described in the BEA-Large study \citep{bea_large}. Dialogues were first segmented into 30-second chunks. Within each segment, speaker changes were explicitly annotated using a dedicated token (\texttt{<sc>}), inspired by Serialized Output Training (SOT) \citep{SOT}. These speaker-change tokens were also included during training; consequently, predicting speaker change markers constituted an explicit learning objective for all fine-tuned models.

Model training was carried out using the \textit{NVIDIA NeMo} framework \citep{nemo}. To ensure comparability with prior work, we adhered to the same experimental setup and fine-tuned an English Fast Conformer~\citep{fastconformer} Large CTC model\footnote{\url{https://huggingface.co/nvidia/stt_en_fastconformer_ctc_large}} across all configurations. All experiments were conducted on a single \textit{NVIDIA RTX 5000 Ada Generation} GPU using a batch size of 16, a learning rate of $5\times10^{-4}$, and a cosine annealing learning-rate scheduler.

To contextualize our results, we report a diverse set of baselines differing in both data augmentation level and strategy. First, we include results for Whisper-v3\footnote{\url{https://huggingface.co/openai/whisper-large-v3}} and the best-performing Fast Conformer Large model (denoted as \texttt{fc\_base} in Table~\ref{tab:best_results}) from \citet{bea_large}. The Whisper model was evaluated without any fine-tuning, whereas the Fast Conformer model was fine-tuned exclusively on the BEA-Dialogue training set.

Secondly, we evaluate two approaches in which the BEA-Dialogue training set was augmented with simulated data constructed without relying on extracted conversational statistics:
\begin{itemize}
    \item \texttt{fc\_nosim}: Utterances selected for simulation were presented independently, without concatenation and consequently without the \texttt{<sc>} token. As a result, the model was trained on generally shorter segments over a larger number of iterations. The \texttt{<sc>} tokens were retained in the original BEA-Dialogue training data, ensuring that the model remained capable of representing speaker change events.
    \item \texttt{fc\_naivesim}: Fixed pauses of 0.25 seconds were inserted between consecutive utterances, instead of modeling pause durations from learned distributions.
\end{itemize}

Finally, we report results obtained using simulated conversations (SC) \citep{Landini2022} generated separately from each of the three corpora. This approach leverages corpus-specific conversational statistics, but differs from our method in that it relies on histogram-based modeling instead of kernel density estimation (KDE) and does not incorporate speaker-aware modeling.

\subsection{Evaluation metrics}
In conversational speech transcription, models may correctly recognize the spoken content while incorrectly ordering utterances based on speakers. For example, when the utterances \textit{A: "Okay."} and \textit{B: "Great."} partially overlap, a system may legitimately output either \textit{"Okay. \texttt{<sc>} Great."} or \textit{"Great. \texttt{<sc>} Okay."}. Conventional evaluation metrics such as word error rate (WER) and character error rate (CER) treat only a single reference ordering as correct, even though the recognized lexical content is accurate in both cases. 

To account for this ambiguity, we additionally report concatenated minimum-permutation error rates, cpWER and cpCER. These metrics compute the error rate over all possible permutations of utterance orderings and retain the minimum value, thereby providing a more faithful assessment of transcription quality in the presence of overlapping speech. In addition, we report \texttt{scAcc}, which measures the percentage of segments for which the model correctly predicts the number of speaker changes.

\subsection{Results}
Results are reported jointly on the development (dev) and evaluation (eval) sets in Table~\ref{tab:best_results}, "Pairs" denotes the number of speaker pairings in which a given speaker appears, and "Method" indicates whether duration-conditioned statistics were applied. Lower WER, CER, cpWER, and cpCER values and higher \texttt{scAcc} indicate better performance.

\begin{table}[t]
\centering
\setlength{\tabcolsep}{4pt}
\resizebox{\textwidth}{!}{
\begin{tabular}{lcc|ccccc|ccccc}
\toprule
\multirow{2}{*}{Approach} &
\multirow{2}{*}{Method} &
\multirow{2}{*}{Pairs} &
\multicolumn{5}{c|}{Dev} &
\multicolumn{5}{c}{Eval} \\
\cmidrule(lr){4-8} \cmidrule(lr){9-13}
 &  &  &
WER & CER & cpWER & cpCER & scAcc &
WER & CER & cpWER & cpCER & scAcc \\
\midrule
Whisper v3 & -- & -- 
& 21.19 & 12.74 & 21.04 & 12.56 & --
& 22.21 & 12.27 & 22.13 & 12.18 & -- \\
fc\_base & -- & --
& 20.07 & 8.27 & 19.86 & 8.09 & \textbf{69.67}
& 21.10 & 9.42 & 21.01 & 9.33 & 81.95 \\
\midrule
fc\_nosim & -- & --
& 17.52 & 7.79 & 17.34 & 7.65 & 67.07
& 18.41 & 8.70 & 18.31 & 8.59 & 78.67 \\
fc\_naivesim & -- & 2
& 17.34 & 7.25 & 17.11 & 7.03 & 68.46
& 18.34 & 8.37 & 18.24 & 8.27 & 82.58 \\
\midrule
BEA & SC & 2
& 16.93 & 7.28 & 16.76 & 7.15 & 67.96
& 17.84 & 8.37 & 17.75 & 8.29 & 81.44 \\
CH & SC & 2
& 17.13 & 7.36 & 16.93 & 7.21 & 69.26
& 18.48 & 8.60 & 18.37 & 8.51 & 81.69 \\
GRASS & SC & 2
& 16.58 & 7.19 & 16.46 & 7.02 & 69.26
& 17.78 & 8.27 & 17.67 & 8.18 & 82.66 \\
\midrule
BEA & SASC & 4
& 16.44 & 7.05 & 16.25 & 6.91 & 69.32
& 17.71 & 8.20 & 17.64 & 8.13 & 82.21 \\
BEA & C-SASC & 3
& \textbf{16.26} & 7.01 & \textbf{16.08} & 6.85 & 68.11
& \textbf{17.51} & 8.18 & \textbf{17.40} & \textbf{8.09} & 82.64 \\
CH & SASC & 3
& 16.45 & 7.02 & 16.23 & \textbf{6.82} & 68.11
& 17.86 & 8.25 & 17.76 & 8.16 & 82.85 \\
CH & C-SASC & 3
& 17.00 & 7.18 & 16.76 & 6.98 & 66.72
& 17.99 & 8.24 & 17.87 & 8.14 & 82.53 \\
GRASS & SASC & 3
& 16.43 & \textbf{6.99} & 16.21 & 6.83 & 68.80
& 17.73 & 8.20 & 17.63 & 8.11 & \textbf{83.01} \\
GRASS & C-SASC & 3
& 16.28 & 7.07 & 16.09 & 6.89 & 67.94
& 17.82 & \textbf{8.16} & 17.74 & \textbf{8.09} & 82.42 \\
\bottomrule
\end{tabular}}
\caption{Best-performing approaches on the Hungarian BEA-Dialogue corpus (development and evaluation sets).}
\label{tab:best_results}
\end{table}

Table~\ref{tab:best_results} shows that even the baseline simulation strategies yield substantial improvements over the non-simulated models, indicating that exposing the system to multi-utterance conversational structure -- even via naive concatenation -- is beneficial. Incorporating corpus-derived conversational statistics through SC leads to improvements in word-level metrics (except for CH on the evaluation set). However, SC degrades character-level performance for both BEA and CH, an effect that is reversed when GRASS-based statistics are used.

SASC and C-SASC yield consistent further improvements across all four transcription metrics and on both the development and evaluation sets when compared to the respective SC-based models.

To assess statistical significance, we apply a bootstrap-based significance testing procedure \citep{Confidence_Intervals} on the evaluation set with a significance level of $\alpha = 0.05$, yielding the following results:
\begin{itemize}
    \item \textit{BEA-based statistics}:
    The C-SASC model shows significant improvements over SC in character-level metrics, but not in word-level metrics. However, significant word-level improvements are observed relative to \texttt{fc\_naivesim}. Notably, SC itself does not outperform significantly the \texttt{fc\_naivesim} model in word-level metrics, whereas the C-SASC model does.
    \item \textit{CH-based statistics}:
    The SASC model shows statistically significant improvements over the corresponding SC model across all word-level and character-level metrics.
    \item \textit{GRASS-based statistics}:
    The C-SASC model does not differ significantly from the corresponding SC model in either word-level or character-level metrics on the evaluation set.
\end{itemize}

Among the unconditional SASC configurations, performance on the dev set is largely similar across corpora, with only minor differences, most notably in cpCER. On the eval set, the differences become more pronounced: models based on BEA and GRASS statistics remain comparable, while the CallHome-based model performs slightly worse. In contrast, C-SASC exhibits greater separation across corpora on the eval set. The BEA-based C-SASC model achieves the strongest overall performance, whereas the CallHome-based variant consistently underperforms. For BEA, C-SASC outperforms SASC across all four metrics on both dev and eval sets. For GRASS, the relative ordering depends on the metric and split: word-level metrics favor C-SASC on the dev set but SASC on the eval set, while character-level metrics exhibit the opposite trend. For CallHome, SASC outperforms C-SASC across all metrics on the dev set and remains superior on word-level metrics on the eval set, with C-SASC showing only marginal advantages on character-level metrics. The weaker performance of CallHome-based C-SASC is consistent with the utterance duration mismatch observed in Fig.~\ref{fig:utterance_length}, where CallHome exhibits the largest deviation from the duration distribution of the simulated material.

The \texttt{scAcc} results do not show clear trends, with the new models all performing slightly worse on the dev and slightly better on the eval set, with no significant differences in this metric.

\begin{figure}[h]
    \centering
    \begin{subfigure}{0.48\linewidth}
        \centering
        \includegraphics[width=\linewidth]{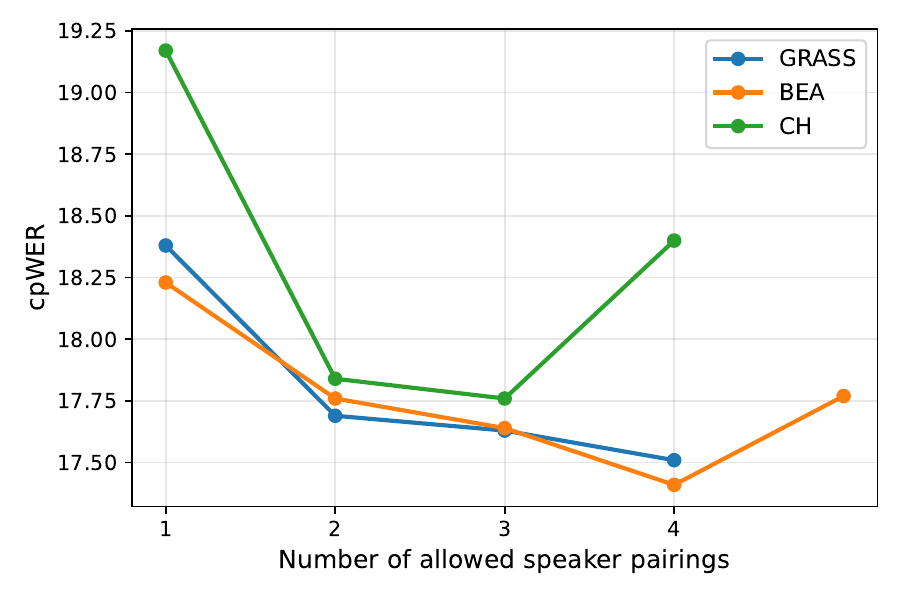}
        \caption{cpWER for SASC as a function of the number of pairs}
    \end{subfigure}
    \hfill
    \begin{subfigure}{0.48\linewidth}
        \centering
        \includegraphics[width=\linewidth]{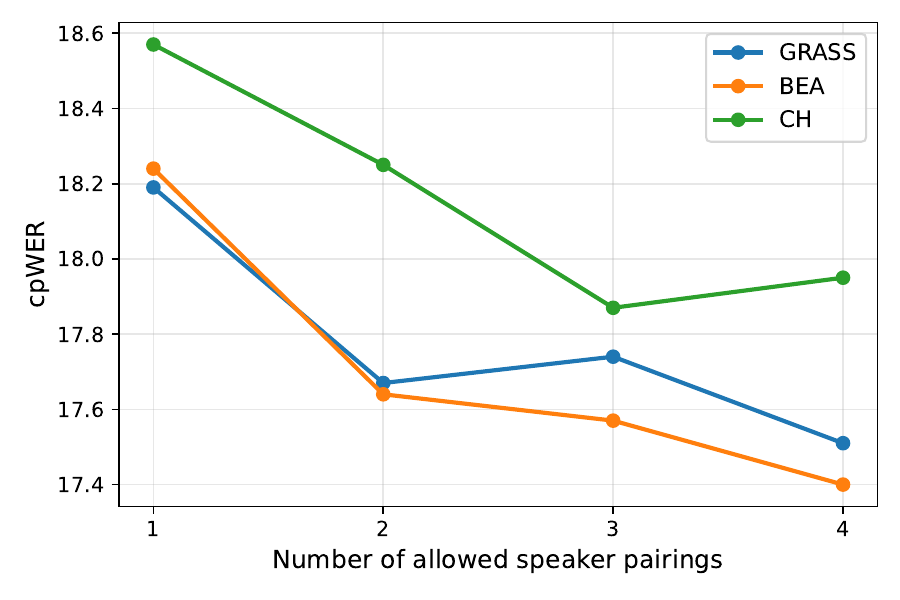}
        \caption{cpWER for C-SASC as a function of the number of pairs}
    \end{subfigure}
    \vspace{0.5em}
    \begin{subfigure}{0.48\linewidth}
        \centering
        \includegraphics[width=\linewidth]{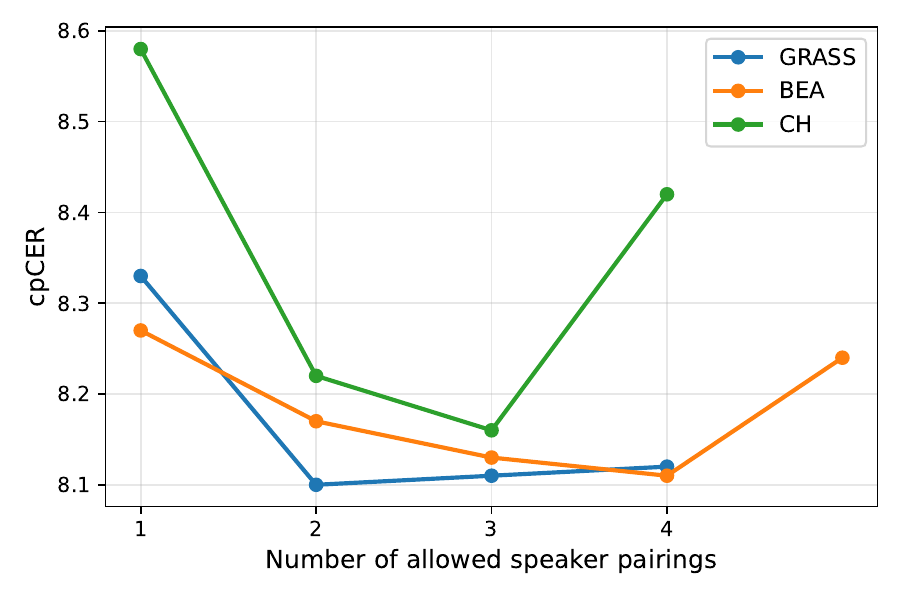}
        \caption{cpCER for SASC as a function of the number of pairs}
    \end{subfigure}
    \hfill
    \begin{subfigure}{0.48\linewidth}
        \centering
        \includegraphics[width=\linewidth]{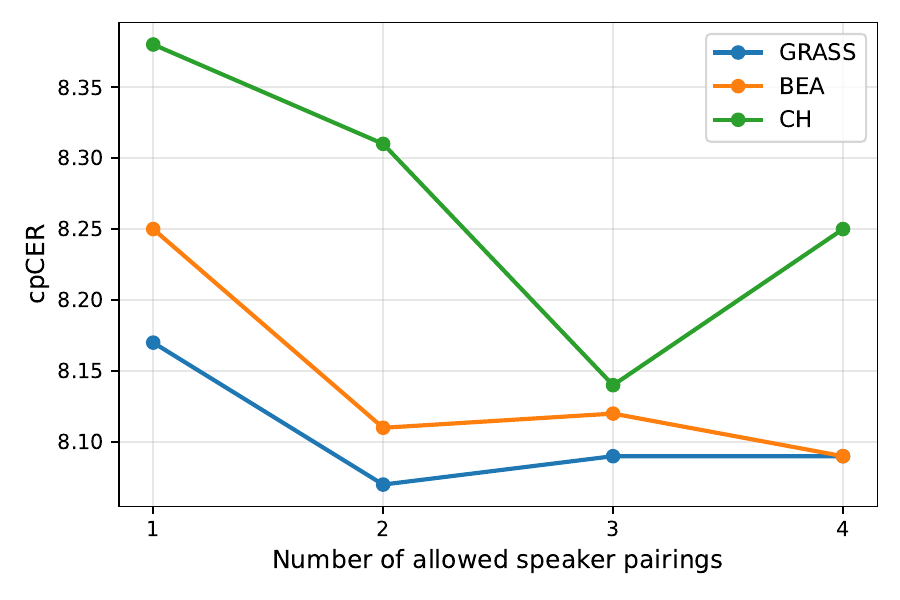}
        \caption{cpCER for C-SASC as a function of the number of pairs}
    \end{subfigure}
    \caption{Effect of simulated dataset size on cpWER and cpCER.}
    \label{fig:pairing_metrics}
\end{figure}

Figure~\ref{fig:pairing_metrics} illustrates the effect of increasing the size of the simulated dataset on cpWER and cpCER (eval set). Across all configurations, models using CallHome-derived statistics consistently perform worst, indicating a poorer fit to the target domain. BEA- and GRASS-based models alternate in relative ranking depending on the metric and simulation size. For cpWER, the best performance is typically achieved with three or four speaker pairings, whereas for cpCER, two or three pairings appear optimal.

Table~\ref{tab:relative_gain} reports the relative gain of C-SASC over SASC on the eval set, with negative values indicating improvement due to conditioning. Overall differences are small, suggesting that for a downstream task such as speech recognition, introducing duration-conditioned pause modeling yields modest gains. Notably, the effect is more consistent for character-level metrics: C-SASC outperforms SASC in 11 out of 12 cpCER configurations, whereas for cpWER the advantage is less systematic, with C-SASC winning in 7 cases, one tie, and SASC performing better in 4 cases. CallHome-based models exhibit the largest variability, consistent with their overall weaker performance, while BEA-based models show the smallest differences between conditional and unconditional variants.

\begin{table}[t]
\centering
\resizebox{\linewidth}{!}{
\begin{tabular}{lcccccccc}
\toprule
\multirow{2}{*}{Approach} 
& \multicolumn{4}{c}{$\Delta$ cpWER (\%)} 
& \multicolumn{4}{c}{$\Delta$ cpCER (\%)} \\
\cmidrule(lr){2-5} \cmidrule(lr){6-9}
& 1 pair & 2 pairs & 3 pairs & 4 pairs
& 1 pair & 2 pairs & 3 pairs & 4 pairs \\
\midrule
BEA   &  0.05 & -0.68 & -0.40 & -0.06 & -0.24 & -0.73 & -0.12 & -0.25 \\
CH    & -3.13 &  2.30 &  0.62 & -2.45 & -2.33 &  1.09 & -0.25 & -2.02 \\
GRASS & -1.03 & -0.11 &  0.62 &  0.00 & -1.92 & -0.37 & -0.25 & -0.37 \\
\bottomrule
\end{tabular}
}
\caption{Relative gain (\%) of conditional (C-SASC) models over unconditional (SASC) models on the evaluation set. Negative values indicate improvement due to conditioning.}
\label{tab:relative_gain}
\end{table}

Although RIR simulation is a default component of the original SASC framework, we exclude it from the main experimental setup, as it generally degrades performance in our setting. Table~\ref{tab:rir_delta} summarizes the relative impact of RIR augmentation on the evaluation set in terms of cpWER and cpCER. Following the LibriConvo methodology~\citep{Libriconvo}, RIRs were applied to 40\% of the simulated conversations.

Across corpora and both SASC variants, RIR augmentation mostly leads to increased error rates. Consistent degradations are observed for CH, while effects on BEA and GRASS are small and mixed. The only notable improvements occur for the BEA-based C-SASC model in cpWER and for the GRASS-based C-SASC model in cpCER; however, these gains are limited and not consistent across metrics.

Overall, the results indicate that RIR augmentation is not beneficial for the majority of configurations considered. We attribute this to a mismatch between the simulated reverberation conditions and the controlled, high-quality studio recordings used for evaluation, where RIR-based augmentation may introduce distortions that are not representative of the target domain.

\begin{table}[t]
\centering
\begin{tabular}{lcc|cc}
\toprule
& \multicolumn{2}{c}{\textbf{SASC}} & \multicolumn{2}{c}{\textbf{C-SASC}} \\
\cmidrule(lr){2-3} \cmidrule(lr){4-5}
\textbf{Corpus} & $\Delta$cpWER (\%) & $\Delta$cpCER (\%) 
                & $\Delta$cpWER (\%) & $\Delta$cpCER (\%) \\
\midrule
BEA   &  0.86 &  1.02 & -1.11 &  0.29 \\
CH    &  2.33 &  3.38 &  3.19 &  3.25 \\
GRASS &  0.86 &  0.74 &  0.25 & -0.62 \\
\bottomrule
\end{tabular}
\caption{Relative performance change (\%) on the evaluation set when enabling RIR augmentation. Positive values indicate degradation.}
\label{tab:rir_delta}
\end{table}

\section{Discussion}
This work investigated speaker-aware conversational data simulation as a means to improve multi-speaker speech recognition, with a particular focus on realistic modeling of pauses, overlaps, and speaker turn-taking. The proposed C-SASC framework extends the original SASC approach by introducing a lightweight but principled conditioning of pause behavior on utterance duration, while preserving the simplicity and efficiency of the original formulation.

Across all experimental settings, conversational simulation proved beneficial. Even a naive simulation strategy (fc\_naivesim in Table~\ref{tab:best_results}) that merely concatenate utterances into dialogues led to improvements over models both trained without simulated conversations and trained on the same data without concatenation. This highlights that exposure to conversational structure itself plays an important role in improving robustness for dialogue transcription tasks. Building on this, both SASC and C-SASC consistently and significantly outperformed the naive baseline approaches, demonstrating that explicitly modeling conversational statistics yields further gains beyond naive data augmentation. They also consistently outperformed a previous simalated conversation approach (SC), varying in significance level.

The comparison between SASC and C-SASC reveals a more nuanced picture. While C-SASC was designed to better capture dependencies between utterance duration and pause behavior, the resulting improvements on downstream ASR performance are generally modest. On average, duration conditioning led to small but consistent gains in character-level metrics (cpCER), whereas improvements in word-level metrics (cpWER) were less systematic. This suggests that more accurate temporal modeling may primarily benefit fine-grained alignment and segmentation decisions -- such as character boundaries and short overlaps -- rather than higher-level lexical recognition.

Importantly, the effectiveness of C-SASC depends on how well the duration statistics of the source corpus match those of the simulated material. The CallHome-based models, which exhibit the largest mismatch in utterance duration distributions relative to the target simulation data, benefited the least from duration conditioning and in some cases performed worse than their unconditional counterparts. In contrast, BEA-based statistics, which are closely aligned with the target domain, consistently yielded stronger results. This finding underscores that conditional modeling increases sensitivity to corpus mismatch and should therefore be applied with care when transferring statistics across domains or languages.

Finally, analysis of simulated dataset size shows diminishing returns beyond a moderate number of reuse of speakers in speaker pairings. While increasing the number of simulated dialogues generally improves performance, optimal results are typically achieved with two to four pairings per speaker. This suggests that diversity in conversational context is beneficial, but excessive reuse of speakers across many simulated dialogues may introduce redundancy rather than additional useful variability.

\section{Conclusion}
In this paper, we evaluated speaker-aware simulated conversations on automatic speech recognition tasks. Beside the comprehensive evaluation of the original SASC method, we presented C-SASC, a duration-conditioned extension. C-SASC augments speaker-dependent pause modeling by conditioning local pause deviations on utterance duration, enabling more realistic modeling of conversational timing while retaining a simple and efficient generative structure.

We evaluated SASC and C-SASC across multiple configurations using statistics extracted from the CallHome, BEA-Dialogue, and GRASS corpora. Extensive experiments on conversational ASR demonstrate that simulation-based data augmentation substantially improves transcription performance, and that incorporating corpus-derived conversational statistics yields consistent gains over naive simulation strategies. While the additional conditioning introduced by C-SASC results in relatively small improvements on downstream ASR metrics, these gains are systematic for character-level error rates and are most pronounced when the source statistics closely match the target domain.

Taken together, our findings suggest that realistic modeling of conversational timing is an important component of dialogue simulation, but that increasingly detailed temporal conditioning offers diminishing returns for standard ASR objectives. Nevertheless, the proposed framework remains attractive due to its interpretability, low computational overhead, and flexibility across languages and recording conditions.

Several directions for future work emerge from the findings of this study. While this work focused on improving downstream ASR performance, more detailed temporal modeling may be particularly valuable for tasks beyond standard transcription. Applications such as speaker diarization, overlap detection, turn-taking prediction, and conversational analysis may benefit more directly from accurate modeling of pause--duration dependencies. Evaluating C-SASC in these settings could provide a more comprehensive assessment of its practical impact. 

Additionally, the proposed framework could be extended beyond real single-speaker recordings to incorporate text-to-speech-generated synthetic speech. In this setting, written dialogues (either from literature or generated by large language models) could be used to generate coherent multi-turn dialogues with explicit control over discourse structure, speaker roles, and interaction patterns, while the SASC-style timing model governs pauses, overlaps, and turn-taking behavior. Such a combination would open new avenues for ensuring linguistic and pragmatic consistency across simulated conversations, allowing fine-grained alignment between textual content, speaker intent, and temporal dynamics. This direction could significantly enhance the scalability and diversity of conversational data generation, while maintaining a principled connection to realistic interaction timing.

Overall, this work represents a step toward more realistic and controllable simulation of conversational speech. Future extensions that balance modeling fidelity with robustness and simplicity may further enhance the utility of simulated dialogues for speech technology development across languages and domains.

\section*{Acknowledgment}
Project No. 2025-2.1.2-EKÖP-KDP-2025-00005 has been implemented with the support provided by the Ministry of Culture and Innovation of Hungary from the National Research, Development and Innovation (NRDI) Fund, financed under the EKÖP\_KDP-25-1-BME-21 funding scheme.

The work was also partially supported by the Hungarian NRDI Fund through the projects NKFIH K143075 and K135038, NKFIH-828- 2/2021(MILAB).


\bibliography{sn-bibliography}

@inproceedings{CHiME-6,
  title     = {CHiME-6 Challenge: Tackling Multispeaker Speech Recognition for Unsegmented Recordings},
  author    = {Shinji Watanabe and Michael Mandel and Jon Barker and Emmanuel Vincent and Ashish Arora and Xuankai Chang and Sanjeev Khudanpur and Vimal Manohar and Daniel Povey and Desh Raj and David Snyder and Aswin Shanmugam Subramanian and Jan Trmal and Bar Ben Yair and Christoph Boeddeker and Zhaoheng Ni and Yusuke Fujita and Shota Horiguchi and Naoyuki Kanda and Takuya Yoshioka and Neville Ryant},
  year      = {2020},
  booktitle = {6th International Workshop on Speech Processing in Everyday Environments (CHiME 2020)},
  pages     = {1--7},
  doi       = {10.21437/CHiME.2020-1},
}

@inproceedings{Landini2022,
  title={From Simulated Mixtures to Simulated Conversations as Training Data for End-to-End Neural Diarization},
  author={Federico Landini and Alicia Lozano-Diez and Mireia D{\'i}ez and Luk{\'a}š Burget},
  booktitle={Interspeech},
  year={2022},
  url={https://api.semanticscholar.org/CorpusID:247939646}
}

@article{Landini2022MultiSpeakerEEND,
  title={Multi-Speaker and Wide-Band Simulated Conversations as Training Data for End-to-End Neural Diarization},
  author={Federico Landini and Mireia D{\'i}ez and Alicia Lozano-Diez and Luk{\'a}š Burget},
  journal={ICASSP 2023},
  year={2022},
  pages={1-5},
  url={https://api.semanticscholar.org/CorpusID:253510723}
}

@inproceedings{Yamashita2022Naturalness,
  title={Improving the Naturalness of Simulated Conversations for End-to-End Neural Diarization},
  author={Natsuo Yamashita and Shota Horiguchi and Takeshi Homma},
  booktitle={The Speaker and Language Recognition Workshop},
  year={2022},
  url={https://api.semanticscholar.org/CorpusID:248377558}
}

@inproceedings{Fujita2019,
  title={End-to-End Neural Speaker Diarization with Permutation-Free Objectives},
  author={Yusuke Fujita and Naoyuki Kanda and Shota Horiguchi and Kenji Nagamatsu and Shinji Watanabe},
  booktitle={Interspeech},
  year={2019},
  url={https://api.semanticscholar.org/CorpusID:202572807}
}

@inproceedings{SOT,
  title={Serialized Output Training for End-to-End Overlapped Speech Recognition},
  author={Naoyuki Kanda and Yashesh Gaur and Xiaofei Wang and Zhong Meng and Takuya Yoshioka},
  booktitle={Interspeech},
  year={2020},
  url={https://api.semanticscholar.org/CorpusID:214714409}
}

@article{Yu2016PIT,
  title={Permutation invariant training of deep models for speaker-independent multi-talker speech separation},
  author={Dong Yu and Morten Kolb{\ae}k and Zheng-Hua Tan and Jesper H{\o}jvang Jensen},
  journal={ICASSP 2017},
  year={2016},
  pages={241-245},
  url={https://api.semanticscholar.org/CorpusID:7331600}
}

@misc{SASC,
      title={From Independence to Interaction: Speaker-Aware Simulation of Multi-Speaker Conversational Timing}, 
      author={Máté Gedeon and Péter Mihajlik},
      year={2025},
      eprint={2509.15808},
      archivePrefix={arXiv},
      primaryClass={cs.SD},
      url={https://arxiv.org/abs/2509.15808}, 
      note={Accepted to ICASSP 2026}
}

@misc{Libriconvo,
      title={LibriConvo: Simulating Conversations from Read Literature for ASR and Diarization}, 
      author={Máté Gedeon and Péter Mihajlik},
      year={2025},
      eprint={2510.23320},
      archivePrefix={arXiv},
      primaryClass={eess.AS},
      url={https://arxiv.org/abs/2510.23320}, 
}

@inproceedings{bartelds2023,
    title = "Making More of Little Data: Improving Low-Resource Automatic Speech Recognition Using Data Augmentation",
    author = "Bartelds, Martijn  and
      San, Nay  and
      McDonnell, Bradley  and
      Jurafsky, Dan  and
      Wieling, Martijn",
    editor = "Rogers, Anna  and
      Boyd-Graber, Jordan  and
      Okazaki, Naoaki",
    booktitle = "Proceedings of the 61st Annual Meeting of the Association for Computational Linguistics (Volume 1: Long Papers)",
    month = jul,
    year = "2023",
    address = "Toronto, Canada",
    publisher = "Association for Computational Linguistics",
    url = "https://aclanthology.org/2023.acl-long.42/",
    doi = "10.18653/v1/2023.acl-long.42",
    pages = "715--729",
}

@INPROCEEDINGS{yang2023,
  author={Yang, Muqiao and Kanda, Naoyuki and Wang, Xiaofei and Wu, Jian and Sivasankaran, Sunit and Chen, Zhuo and Li, Jinyu and Yoshioka, Takuya},
  booktitle={ICASSP 2023 - 2023 IEEE International Conference on Acoustics, Speech and Signal Processing (ICASSP)}, 
  title={Simulating Realistic Speech Overlaps Improves Multi-Talker ASR}, 
  year={2023},
  volume={},
  number={},
  pages={1-5},
  doi={10.1109/ICASSP49357.2023.10094928}}

@inproceedings{libriheavymix,
  title     = {{LibriheavyMix: A 20,000-Hour Dataset for Single-Channel Reverberant Multi-Talker Speech Separation, ASR and Speaker Diarization}},
  author    = {Zengrui Jin and Yifan Yang and Mohan Shi and Wei Kang and Xiaoyu Yang and Zengwei Yao and Fangjun Kuang and Liyong Guo and Lingwei Meng and Long Lin and Yong Xu and Shi-Xiong Zhang and Daniel Povey},
  year      = {2024},
  booktitle = {{Interspeech 2024}},
  pages     = {702--706},
  doi       = {10.21437/Interspeech.2024-90},
  issn      = {2958-1796},
}

@misc{bea_large,
      title={{Toward Conversational Hungarian Speech Recognition: Introducing the BEA-Large and BEA-Dialogue Datasets}}, 
      author={Máté Gedeon and Piroska Zsófia Barta and Péter Mihajlik and Tekla Etelka Gráczi and Anna Kohári and Katalin Mády},
      year={2025},
      eprint={2511.13529},
      archivePrefix={arXiv},
      primaryClass={cs.CL},
      url={https://arxiv.org/abs/2511.13529}, 
}

@article{speech_planning,
    author = {{Krivokapi{\v{c}}, Jelena and Styler, Will and Byrd, Dani}},
    title = {The role of speech planning in the articulation of pauses},
    journal = {The Journal of the Acoustical Society of America},
    volume = {151},
    number = {1},
    pages = {402-413},
    year = {2022},
    month = {01},
    issn = {0001-4966},
    doi = {10.1121/10.0009279},
    url = {https://doi.org/10.1121/10.0009279}
}

@article{nadaraya,
author = {Nadaraya, E. A.},
title = {On Estimating Regression},
journal = {Theory of Probability \& Its Applications},
volume = {9},
number = {1},
pages = {141-142},
year = {1964},
doi = {10.1137/1109020},
URL = {https://doi.org/10.1137/1109020},
eprint = {https://doi.org/10.1137/1109020},
}

@article{watson,
 ISSN = {0581572X},
 URL = {http://www.jstor.org/stable/25049340},

 author = {Geoffrey S. Watson},
 journal = {Sankhyā: The Indian Journal of Statistics, Series A (1961-2002)},
 number = {4},
 pages = {359--372},
 publisher = {Springer},
 title = {Smooth Regression Analysis},
 urldate = {2026-01-07},
 volume = {26},
 year = {1964}
}

@article{Yeo-Johnson,
 ISSN = {00063444, 14643510},
 URL = {http://www.jstor.org/stable/2673623},
 author = {In-Kwon Yeo and Richard A. Johnson},
 journal = {Biometrika},
 number = {4},
 pages = {954--959},
 publisher = {[Oxford University Press, Biometrika Trust]},
 title = {A New Family of Power Transformations to Improve Normality or Symmetry},
 urldate = {2025-09-04},
 volume = {87},
 year = {2000}
}

@book{silverman1986,
  title={Density Estimation for Statistics and Data Analysis},
  author={Silverman, B. W.},
  year={1986},
  publisher={Chapman \& Hall},
  address={London},
  doi={10.1007/978-1-4899-3324-9},
  series={Monographs on Statistics and Applied Probability}
}

@book{scott1992,
  title={Multivariate Density Estimation: Theory, Practice and Visualization},
  author={Scott, D. W.},
  year={1992},
  publisher={John Wiley \& Sons},
  address={New York},
  doi={10.1002/9780470316849},
  series={Wiley Series in Probability and Statistics}
}

@inproceedings{holmes2012,
  title={Fast Nonparametric Conditional Density Estimation},
  author={Michael P. Holmes and Alexander G. Gray and Charles Lee Isbell},
  booktitle={Conference on Uncertainty in Artificial Intelligence},
  year={2007},
  url={https://api.semanticscholar.org/CorpusID:7714554}
}

@misc{CallHome,
  author       = {Canavan, Alexandra and Graff, David and Zipperlen, George},
  title        = {CALLHOME American English Speech},
  howpublished = {Web Download},
  address      = {Philadelphia},
  publisher    = {Linguistic Data Consortium},
  year         = {1997},
  note         = {LDC Catalog No.: LDC97S42, ISBN: 1-58563-111-6, ISLRN: 952-976-147-406-5},
  doi          = {10.35111/exq3-x930},
  url          = {https://catalog.ldc.upenn.edu/LDC97S42}
}

@inproceedings{grass,
    title = "{GRASS}: the Graz corpus of Read And Spontaneous Speech",
    author = "Schuppler, Barbara  and
      Hagmueller, Martin  and
      Morales-Cordovilla, Juan A.  and
      Pessentheiner, Hannes",
    editor = "Calzolari, Nicoletta  and
      Choukri, Khalid  and
      Declerck, Thierry  and
      Loftsson, Hrafn  and
      Maegaard, Bente  and
      Mariani, Joseph  and
      Moreno, Asuncion  and
      Odijk, Jan  and
      Piperidis, Stelios",
    booktitle = "Proceedings of the Ninth International Conference on Language Resources and Evaluation ({LREC}'14)",
    month = may,
    year = "2014",
    address = "Reykjavik, Iceland",
    publisher = "European Language Resources Association (ELRA)",
    url = "https://aclanthology.org/L14-1341/",
    pages = "1465--1470",
}

@misc{Silero,
  author = {Silero Team},
  title = {{Silero VAD: pre-trained enterprise-grade Voice Activity Detector (VAD), Number Detector and Language Classifier}},
  year = {2024},
  publisher = {GitHub},
  journal = {GitHub repository},
  howpublished = {\url{https://github.com/snakers4/silero-vad}},
  commit = {insert_some_commit_here},
  email = {hello@silero.ai}
}

@inproceedings{mfa,
  title     = {{Montreal Forced Aligner: Trainable Text-Speech Alignment Using Kaldi}},
  author    = {Michael McAuliffe and Michaela Socolof and Sarah Mihuc and Michael Wagner and Morgan Sonderegger},
  year      = {2017},
  booktitle = {Interspeech 2017},
  pages     = {498--502},
  doi       = {10.21437/Interspeech.2017-1386},
  issn      = {2958-1796},
}

@software{Confidence_Intervals,
author = {Ferrer, Luciana and Riera, Pablo},
title = {Confidence Intervals for evaluation in machine learning},
url = {https://github.com/luferrer/ConfidenceIntervals}}

@article{Schegloff2000, title={Overlapping talk and the organization of turn-taking for conversation}, volume={29}, DOI={10.1017/S0047404500001019}, number={1}, journal={Language in Society}, author={Schegloff, Emanuel A.}, year={2000}, pages={1–63}}

@article{Heldner2010,
  title={Pauses, gaps and overlaps in conversations},
  author={Mattias Heldner and Jens Edlund},
  journal={J. Phonetics},
  year={2010},
  volume={38},
  pages={555-568},
  url={https://api.semanticscholar.org/CorpusID:7900155}
}

@article{fastconformer,
  title={Fast Conformer With Linearly Scalable Attention For Efficient Speech Recognition},
  author={Dima Rekesh and Samuel Kriman and Somshubra Majumdar and Vahid Noroozi and He Juang and Oleksii Hrinchuk and Ankur Kumar and Boris Ginsburg},
  journal={2023 IEEE Automatic Speech Recognition and Understanding Workshop (ASRU)},
  year={2023},
  pages={1-8},
  url={https://api.semanticscholar.org/CorpusID:258564901}
}

@misc{nemo,
      title={NeMo: a toolkit for building AI applications using Neural Modules}, 
      author={Oleksii Kuchaiev and Jason Li and Huyen Nguyen and Oleksii Hrinchuk and Ryan Leary and Boris Ginsburg and Samuel Kriman and Stanislav Beliaev and Vitaly Lavrukhin and Jack Cook and Patrice Castonguay and Mariya Popova and Jocelyn Huang and Jonathan M. Cohen},
      year={2019},
      eprint={1909.09577},
      archivePrefix={arXiv},
      primaryClass={cs.LG},
      url={https://arxiv.org/abs/1909.09577}, 
}

@article{Mady2024RevisedAnnotation,
  author    = {M{\'a}dy, Katalin and Gr{\'a}czi Tekla Etelka and Koh{\'a}ri, Anna and Mihajlik, P{\'e}ter},
  title     = {Revised annotation conventions in Hungarian speech corpora},
  journal   = {Besz{\'e}dtudom{\'a}ny / Speech Science},
  volume    = {4},
  number    = {1},
  pages     = {185--202},
  year      = {2024},
  note      = {18 p.},
  doi       = {}
}

@InProceedings{bea2014,
author="Neuberger, Tilda
and Gyarmathy, Dorottya
and Gr{\'a}czi, Tekla Etelka
and Horv{\'a}th, Vikt{\'o}ria
and G{\'o}sy, M{\'a}ria
and Beke, Andr{\'a}s",
editor="Sojka, Petr
and Hor{\'a}k, Ale{\v{s}}
and Kope{\v{c}}ek, Ivan
and Pala, Karel",
title="Development of a Large Spontaneous Speech Database of Agglutinative Hungarian Language",
booktitle="Text, Speech and Dialogue",
year="2014",
publisher="Springer International Publishing",
address="Cham",
pages="424--431",
isbn="978-3-319-10816-2"
}

@inproceedings{LibriTTS,
  title     = {LibriTTS: A Corpus Derived from LibriSpeech for Text-to-Speech},
  author    = {Heiga Zen and Viet Dang and Rob Clark and Yu Zhang and Ron J. Weiss and Ye Jia and Zhifeng Chen and Yonghui Wu},
  year      = {2019},
  booktitle = {Interspeech 2019},
  pages     = {1526--1530},
  doi       = {10.21437/Interspeech.2019-2441},
  issn      = {2958-1796},
}

\end{document}